\documentclass{article}

\usepackage{arxiv}

\usepackage[utf8]{inputenc} % allow utf-8 input
\usepackage[T1]{fontenc}    % use 8-bit T1 fonts
\usepackage{hyperref}       % hyperlinks
\usepackage{url}            % simple URL typesetting
\usepackage{booktabs}       % professional-quality tables
\usepackage{amsfonts}       % blackboard math symbols
\usepackage{nicefrac}       % compact symbols for 1/2, etc.
\usepackage{microtype}      % microtypography
\usepackage{lipsum}
\usepackage{graphicx}
\usepackage{amsmath}
\usepackage{amssymb}
\usepackage{xcolor}
\graphicspath{ {./images/} }
\usepackage{hyphenat}
\usepackage{caption}
\usepackage{subcaption}
\usepackage{multirow}
\usepackage{authblk}

\title{Emotion Recognition Through Observer's Physiological Signals}
\author[1,2]{\small Yang Liu}
\author[1]{\small Tom Gedeon}
\author[1]{\small Sabrina Caldwell}
\author[1]{\small Shouxu Lin}
\author[1]{\small Zi Jin}

\affil[1]{\footnotesize Research School of Computer Science, Australian National University, Australia}
\affil[2]{\footnotesize Data61, CSIRO, Australia}
\affil[ ]{\textit {\{yang.liu3, tom.gedeon, sabrina.caldwell, u5898871, zi.jin\}@anu.edu.au}}
\begin{document}
\maketitle

\begin{abstract}
Emotion recognition based on physiological signals is a hot topic and has a wide range of applications, like safe driving, health care and creating a secure society. This paper introduces a physiological dataset PAFEW, which is obtained using movie clips from the Acted Facial Expressions in the Wild (AFEW) dataset~\cite{AFEW} as stimuli. To establish a baseline, we use the electrodermal activity (EDA) signals in this dataset and extract 6 features from each signal series corresponding to each movie clip to recognize 7 emotions, i.e., Anger, Disgust, Fear, Happy, Surprise, Sad and Neutral. Overall, 24 observers participated in our collection of the training set, including 19 observers who participated in only one session watching 80 videos from 7 classes and 5 observers who participated multiple times and watched all the videos. All videos were presented in an order balanced fashion. Leave-one-observer-out was employed in this classification task. We report the classification accuracy of our baseline, a three-layer network, on this initial training set while training with signals from all participants, only single participants and only multiple participants. We also investigate the recognition accuracy of grouping the dataset by arousal or valence, which achieves 68.66\% and 72.72\% separately. 
Finally, we provide a two-step network. The first step is to classify the features into high/low arousal or positive/negative valence by a network.% according to the 2D emotion model
Then the arousal/valence middle output of the first step is concatenated with feature sets as input of the second step for emotion recognition. We found that adding arousal or valence information can help to improve the classification accuracy. In addition, the information of positive/ negative valence boosts the classification accuracy to a higher degree on this dataset.
% add conclusions of experiment results
\end{abstract}

% keywords can be removed

\section{Introduction}
\label{sec:Introduction}
% Introduction to this task
Emotions play a crucial role in human work and life through affecting physiological and psychological status. The long-term accumulation of negative emotions can lead to depression, which has a severe impact on health. Thus, recognizing emotions is a profound research direction, which has a wide range of applications in safe driving~\cite{safe_driving}, health care especially mental health monitoring~\cite{Mental_Health}, creating a secure society~\cite{social_security} and human-computer interaction design.

Emotions often arise spontaneously rather than through conscious effort, which makes recognizing them precisely and in a timely fashion a challenging task. The existing emotion recognizing methods in this area can be classified into two major categories. One is employing human external physical signals, such as facial expressions, gestures, postures, body movements, speech etc. Although these signals are easy to collect, the reliability of data is questionable. This is because these behaviors can be controlled consciously in response to different social situations so they do not always reflect true emotions. The other way is using internal physiological signals, such as electroencephalogram (EEG), skin temperature (SKT), electrocardiogram (ECG), electromyogram (EMG), galvanic skin response/electrodermal activity (GSR/EDA), respiration (RSP) and photoplethysmography (PPG) etc. These physiological signals are sensitive to emotional fluctuations~\cite{EDA} and hard to manipulate. Compared to external physical signals, which almost always communicate identity, appearance and behavior information, physiological signals show an advantage in protecting participants' privacy. In addition, instead of carrying a camera to capture physical signals like facial expressions in strict lab settings, physiological sensors such as E4 wristbands can be more portable and convenient, enabling recording of real-world signals even when the participants are out of the lab for a long period.
These factors make physiological signals more suitable to be used in emotion recognition. This paper adopts physiological signals as the key component of the dataset.

However, while recording data is easy, emotion recognition is still a hard problem because it is time-consuming and often even unfeasible to obtain ground truth human emotion labels. % How to solve the labelling problem? 
Research has been conducted with different carefully selected emotional media as stimuli, such as text, images, audio and videos, to induce corresponding emotion states and collect physiological signals. The labels of the stimuli are treated as ground truth. Sharma et al.~\cite{text} try to recognize stress responses based on text and find that signals like skin temperature and heart rate can achieve an accuracy of 98\%. Picard et al.~\cite{Picard01Image} utilize images to elicit eight emotions. Jang et.al.~\cite{Jang2015Sound} employ sound to arouse three emotions (boredom, pain, surprise) and acquire ECG, EDA ,SKT, and PPG signals for classification. Kim et.al.~\cite{Kim04AudioVisual} combine audio and visual stimuli and record ECG, SKT variation and EDA signals to predict three categories of emotions. %related work of video datasets; limitation
In this paper, clips from classic movies in the AFEW dataset~\cite{AFEW} are employed as stimuli to arouse six basic emotions (anger, disgust, fear, happiness, sadness, surprise) and the neutral class. 
% The label of each clip are used as the ground truth emotion label. % add more related works

%since people resonate when watching great movies
% Most existing approaches are proposed largely for tightly controlled environments.(To be verified) 
% Contribution
The contributions of this paper are twofold. First, we propose a large physiological dataset corresponding to AFEW~\cite{AFEW} for seven basic emotion recognition, which is named as PAFEW. This dataset contains EDA, SKT, PPG, inter beat intervals and heart rate temporal signals, as well as pupil information, covering a range of critical physiological signals for emotion recognition. 
% what's difference between other video stimuli datasets?
Second, we introduce a three-layer network as well as feature extraction as a baseline for this dataset, which achieves 42.08\% on the single participant dataset. We also investigate the accuracy of classifying the dataset by arousal or valence, which achieves 68.66\% and 72.72\% separately. Based on this finding, a two-step network is provided. We classify the features into high/low arousal or positive/negative valence at the first step and the arousal/valence middle output is added to the input feature sets to recognize seven emotions in the second step. We found that adding arousal or valence information can help to improve the recognition accuracy. Furthermore, the information of valence boosts the classification result to a higher degree as far as 4.9\%.
We hope this work is able to facilitate and promote studies in this area.

\section{Methods}
\label{sec:Methods}
\subsection{Data Collection}
\textbf{Experiment Design.} We display movie clips from AFEW~\cite{AFEW} via a website to the participants. The clips contain seven classes of videos with labels: Anger, Disgust, Fear, Happy, Sad, Surprise and Neutral. In total, the 773 videos from the training set, the 383 videos from the validation set and the 653 videos from the test set are employed as stimuli. The length of videos are from 300ms to 5400ms. % need to filter out the videos < 3s ? (How many of them?)
EDA, SKT, PPG, inter beat intervals, heart rate signals and information related to pupils are monitored and collected by Empatica E4 wristband~\cite{E4} and the Eye Tribe~\cite{EyeTribe} gaze tracked. We release the E4 EDA dataset first and the rest along with the complementary Eye Tribe dataset will be released later. In this paper, we only report the collection and results of the training set. More details and results related to the validation and test sets will be updated later. % Why we choose this dataset as stimuli, contribution?

An ideal emotional dataset should gather subjects' genuine emotion elicited by the videos instead of being related to whatever comes to their mind unrelated to the task. The emotion should occurs in subjects' natural state as well as their familiar environment. The subjects should really feel the emotion internally without being influenced by the knowledge of being recorded in an experiment. However, this is unfeasible in reality due to privacy and ethics concerns. In our experiment, all the participants were asked to sign a written consent form before their voluntary participation in the study, which was approved by the Human Research Ethics Committee of The Australian National University (ANU). Furthermore, our participants are clearly aware that the E4 wristband and Eye Tribe are recording their physiological signals during data collection. However, it is easy to forget about a wrist based device and the Eye Tribe is non-intrusive, so the recording will not make much difference. Although the knowledge they were being recorded might affect some observers, it is the state of the art in terms of the balance in recording data ethically. To eliminate the impact caused by other factors, we ask the observers to sit in a familiar University environment with a consistent setting between participants. At the beginning of the experiment, we have a short chat with each participant and explain the experiment. To reduce the influence of their mental state and memories and to avoid subjects being distracted by other factors around, we also ask them to answer some questions right after watching each clip, focusing their attention. The questions are: ‘How does this expression presented in the video look to you?’ (a five-order Likert scale, 1 is completely fake, 5 is completely real), “Have you seen this video before?” (Yes, No; only a very few ‘Yes’ results were obtained) and “How do you rate your confidence level?” (also a five-order Likert scale, 1 is week while 5 is strong) respectively. We record subjects' physiological signals while they are watching videos and answering the first question, which not only accommodates the delay of arousing an emotion but also prevents the emotions from being affected by the following two questions. All of these settings are to facilitate the elicitation of an internal feeling of each emotion by subjects.

\textbf{Participants.} There are 24 subjects who participated in our data collection of the training set in total, including 16 females and 8 males. Among them, there are 20 subjects from China and 4 subjects from other races. The median age is 23 years old and mean age is 24 years old with a standard deviation of 6.16.

The subjects are divided into two different groups: single participants and multiple participants. Single participants are subjects who only participate in one session and view around 80 video clips from all seven classes. Videos with the same emotion are displayed consecutively instead of in a random order so that objects do not need to oscillate between different emotion states. This design also allows observers to have time to become immersed in one kind of emotion, which helps them reliably feel each emotion. The order of videos within an emotion is order balanced, as is the order of presentation of the groups of emotion videos. 
In this way, every 9 participants covers the training set and every 23 participants covers all the videos in AFEW~\cite{AFEW}. 
Since the physiological signals from individuals are noisy and reflect individual differences, the full PAFEW dataset requires 115 participants to cover each video 5 times. The initial results in this paper reflect the initial data collection, which was from 19 single participants who watched the videos from the training set. Multiple participants are subjects who participate in multiple sessions and observe all videos from one class each time. They need to attend seven sessions to collect all their response to the seven emotions. So far, 5 multiple participants are included in our training set. The attributes of our dataset are summarized in Table~\ref{Tab: Attributes of our PAFEW dataset}.

\begin{table*}
\centering
\caption{Attributes of our PAFEW dataset}
\label{Tab: Attributes of our PAFEW dataset}
\begin{tabular}{l|l|c|c}
\hline\hline
\multicolumn{2}{l|}{Attribute}                                   & AFEW       & PAFEW (Our set) \\  \hline \hline
\multicolumn{2}{l|}{Length of sequences}                         & 300-5400 ms & 480-6040 ms             \\ \hline
\multicolumn{2}{l|}{Number of sequences}                         & 1809 & 4023    \\ \hline
\multicolumn{2}{l|}{Number of participants}                      & -- & 24       \\ \hline
\multicolumn{2}{l|}{Maximum number of participants for a clip}   & -- & 9           \\ \hline
\multicolumn{2}{l|}{Minimum number of participants for a clip}   & -- & 2              \\ \hline
\multicolumn{2}{l|}{Mean number of participants for a clip}      & -- &  5.2               \\ \hline
\multirow{7}{*}{Number of sequences per expression} & Anger     & 133  & 716    \\
                                                    & Disgust   & 74 & 323      \\
                                                    & Fear      & 81  & 516     \\
                                                    & Sadness   & 117  & 630    \\
                                                    & Happiness & 150  & 807 \\
                                                    & Surprise  & 74  & 545     \\
                                                    & Neutral   & 144  & 486    \\
\hline\hline
\end{tabular}
\end{table*}

\subsection{Baseline Method}
\textbf{Preprocessing.} The E4 wristband in our experiment collects skin conductance signals at a rate of 4Hz~\cite{E4}. Since subjects watch videos with the same label in consecutive periods, their signals for each category of emotions could be treated as continuous in time for the purpose of normalization. To provide an initial baseline, we only employ EDA signals as our training and validation set. When we plot the EDA sequences for each emotion of each subject, we found that the data obtained from different observers varies significantly (from around 0 to 6.68). We apply min-max normalization to the original data, to reduce between-participant differences. Suppose $\hat{X}_1, ..., \hat{X}_T$ are the original EDA signals for a kind of emotion from a subject, i.e., they correspond to all the videos from the same emotion class, the normalization could be expressed as:
\begin{equation}
    X_i = \frac{\hat{X}_i-\hat{X}_{min}}{\hat{X}_{max}-\hat{X}_{min}},
\end{equation}
where $\hat{X}_{min} = Min\{\hat{X}_1, ..., \hat{X}_T\}$ and $\hat{X}_{max} = Max\{\hat{X}_1, ..., \hat{X}_T\}$.

\begin{figure}
     \centering
     \begin{subfigure} [b]{0.3\textwidth}
         \centering
         \includegraphics[width=\textwidth]{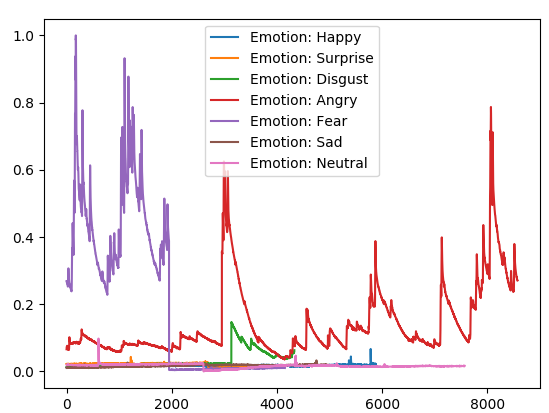}
         \caption{an single participant}
         \label{fig:single participant EDA}
     \end{subfigure}
     \hfill
     \begin{subfigure}[b]{0.3\textwidth}
         \centering
         \includegraphics[width=\textwidth]{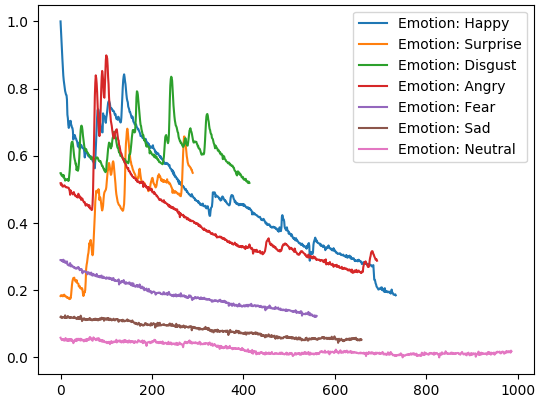}
         \caption{a multiple participant}
         \label{fig:multiple participant EDA}
     \end{subfigure}
     \hfill
     \begin{subfigure}[b]{0.3\textwidth}
         \centering
         \includegraphics[width=\textwidth]{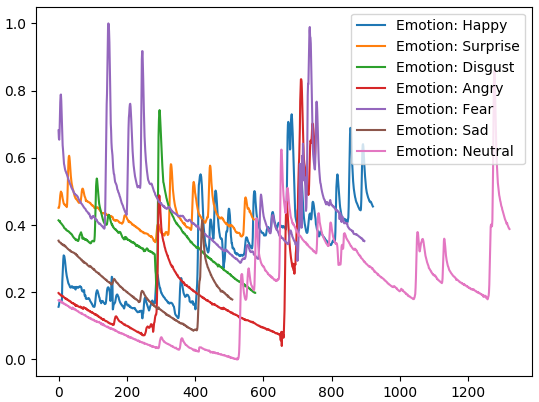}
         \caption{another multiple participant}
         \label{fig:multiple participant EDA}
     \end{subfigure}
     \hfill     
\caption{EDA signals from three participant samples} % Add some conclusions
\label{fig:EDA signal samples}
\end{figure}

After normalization, we plot the signal sequences of each subjects again to check whether there is any obvious abnormal values. Using unobtrusive sensors on people leads to noise and sometimes degraded signals.
% We've met a case of poor contact of devices which degraded the signals to a great extend before. 
Although it is hard for us to tell whether they are good physiological signals from the plots of 1D signal waveform, this process can help to avoid some errors like poor device contact. We show three normalized EDA signal examples in Figure~\ref{fig:EDA signal samples}, which are from one single participant and two multiple participants. These plots illustrate that different subjects respond considerably differently to different emotion stimuli. Some emotion-inducing movie clips may affect some viewers more than others. However, we include these complicated factors in our dataset to better simulate the real world and get models with strong generalization ability. In addition, the movement of subjects may also add noise, so we use an 11-point median filter to remove noise and smooth data sequences.

\textbf{Feature Extraction.} We extract 6 features from the EDA sequences corresponding to each video, which could be summarized to three groups, as listed in Table~\ref{Tab: Feature extraction}. The first group includes 4 basic statistical variables: the maximum, minimum, mean and variance values of the sequence.
The second group contains a statistical variable of the First-Order difference series, the mean of the absolute value of First-Order difference. The First-Order difference is calculated by the standard deviation between two consecutive signals, which approximates a gradient. This variable describes the variation magnitude of the EDA sequence. The last group contains a statistical variable of the Second-Order difference sequence, the mean of the absolute value of Second-Order difference. It illustrates the variation magnitude of the gradient of this series, which could be treated as nonlinear combinations of other features. 
These 6 features cover some typical features reported in the emotion physiology literature~\cite{GSR_features}.
They are easy to compute in an online way, which makes them advantageous in the future for real time recognition.

\begin{table*}
\centering
\caption{Feature extraction}
\label{Tab: Feature extraction}
\begin{tabular}{l|c|c|c}
\hline\hline
Group & Feature & Formula & Description \\
\hline\hline
\multirow{4}{5.5em}{Basic Statistical Variables} & Max 
& $X_{max} = Max\{X_1, ..., X_N\}$ & \multirow{4}{14em}{$X_1, ..., X_N$ are the preprocessed EDA signals corresponding to a video from a subject.} \\
& Min & $X_{min} = Min\{X_1, ..., X_N\}$ & \\
& Mean & $X_{mean} = \frac{1}{N}\sum_{i=1}^{N}X_i$ & \\
& Variance & $X_{var} = \frac{1}{N-1}\sum_{i=1}^{N}(X_i-X_{mean})^2$ & \\ \hline
\multirow{4}{5.5em}{First-Order Difference Statistical Variables} & 
&  & \multirow{4}{14em}{$X_1^{'}, ..., X_N^{'}$ are the First-Order Difference of the preprocessed EDA signals.
$X_i^{'} = X_{i+1}-X_{i}$, \newline where, $i=1, ..., N-1$} \\ & Mean Abs Diff & $X_{mean}^{'} = \frac{1}{N}\sum_{i=1}^{N}|X_i^{'}|$ & \\ & & & \\ & & & \\\hline
\multirow{4}{5.5em}{Second-order Difference Statistical Variables} & 
&  & \multirow{4}{14em}{$X_1^{''}, ..., X_N^{''}$ are the Second-Order Difference of the preprocessed EDA signals.
$X_i^{''} = X_{i+1}^{'}-X_{i}^{'}$ \newline where, $i=1, ..., N-2$} \\ & Mean Second Abs Diff & $X_{mean}^{''} = \frac{1}{N}\sum_{i=1}^{N}|X_i^{''}|$  & \\ & & & \\ & & & \\
\hline \hline
\end{tabular}
\end{table*}

\begin{figure}
     \centering
     \begin{subfigure} [b]{0.155\textwidth}
         \centering
         \includegraphics[width=\textwidth]{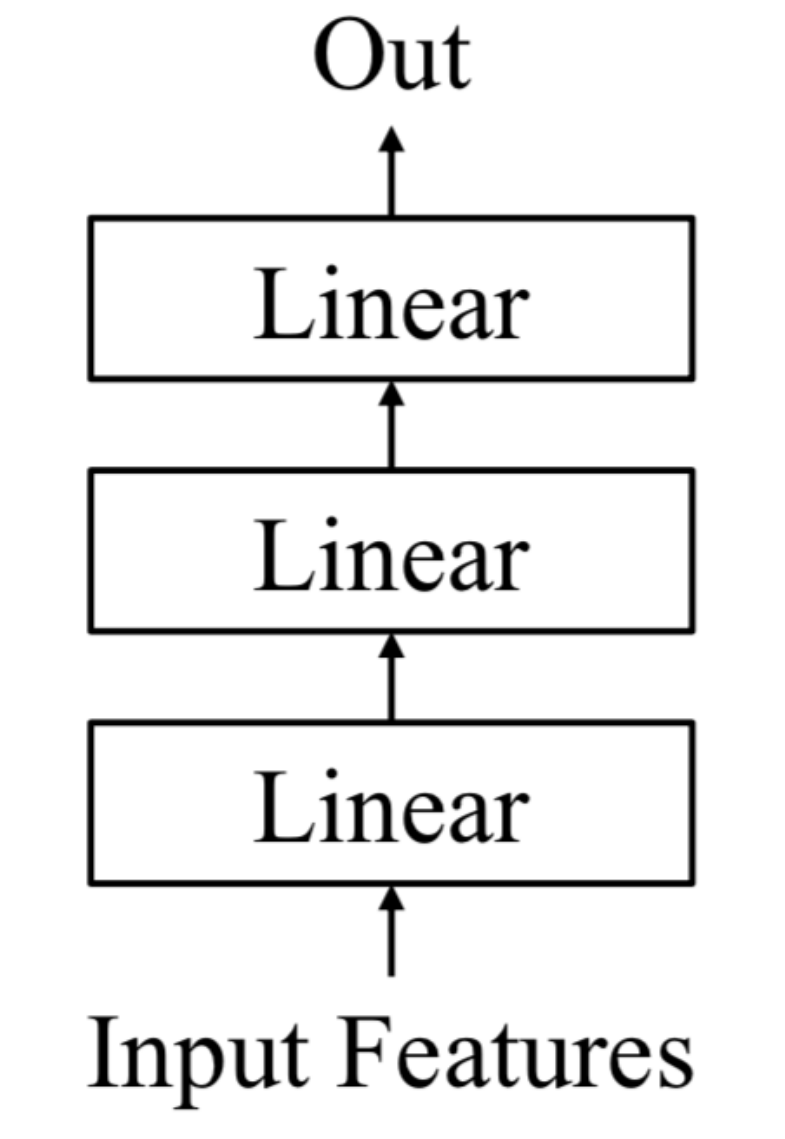}
         \caption{Basic Network Architecture}
         \label{fig:NetArchi1}
     \end{subfigure}
     \hfill
     \begin{subfigure} [b]{0.35\textwidth}
         \centering
         \includegraphics[width=\textwidth]{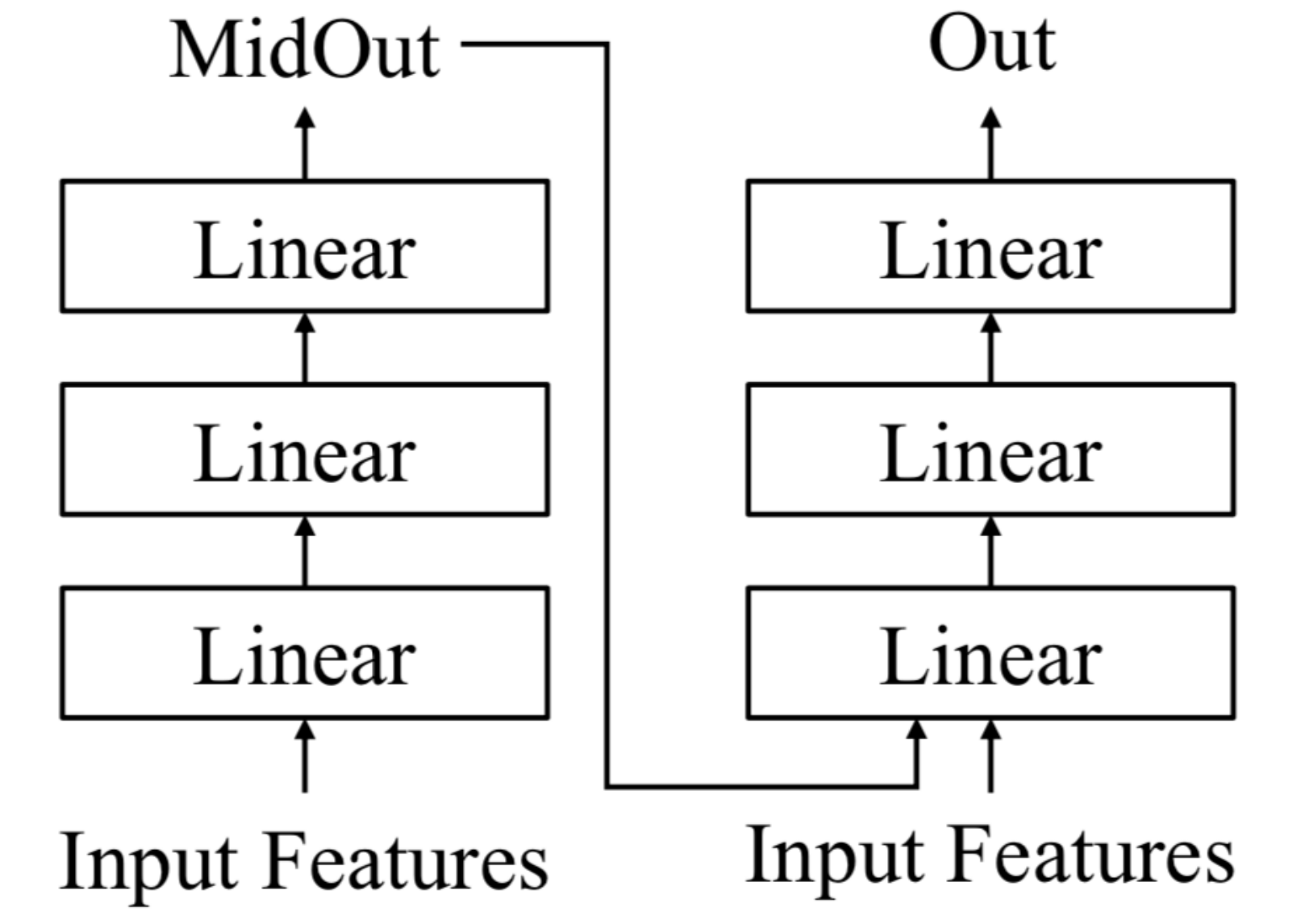}
         \caption{Two-Step Network Architecture}
         \label{fig:NetArchi}
     \end{subfigure}
     \hfill
     \begin{subfigure} [b]{0.3\textwidth}
     \centering
     \includegraphics[width=\textwidth]{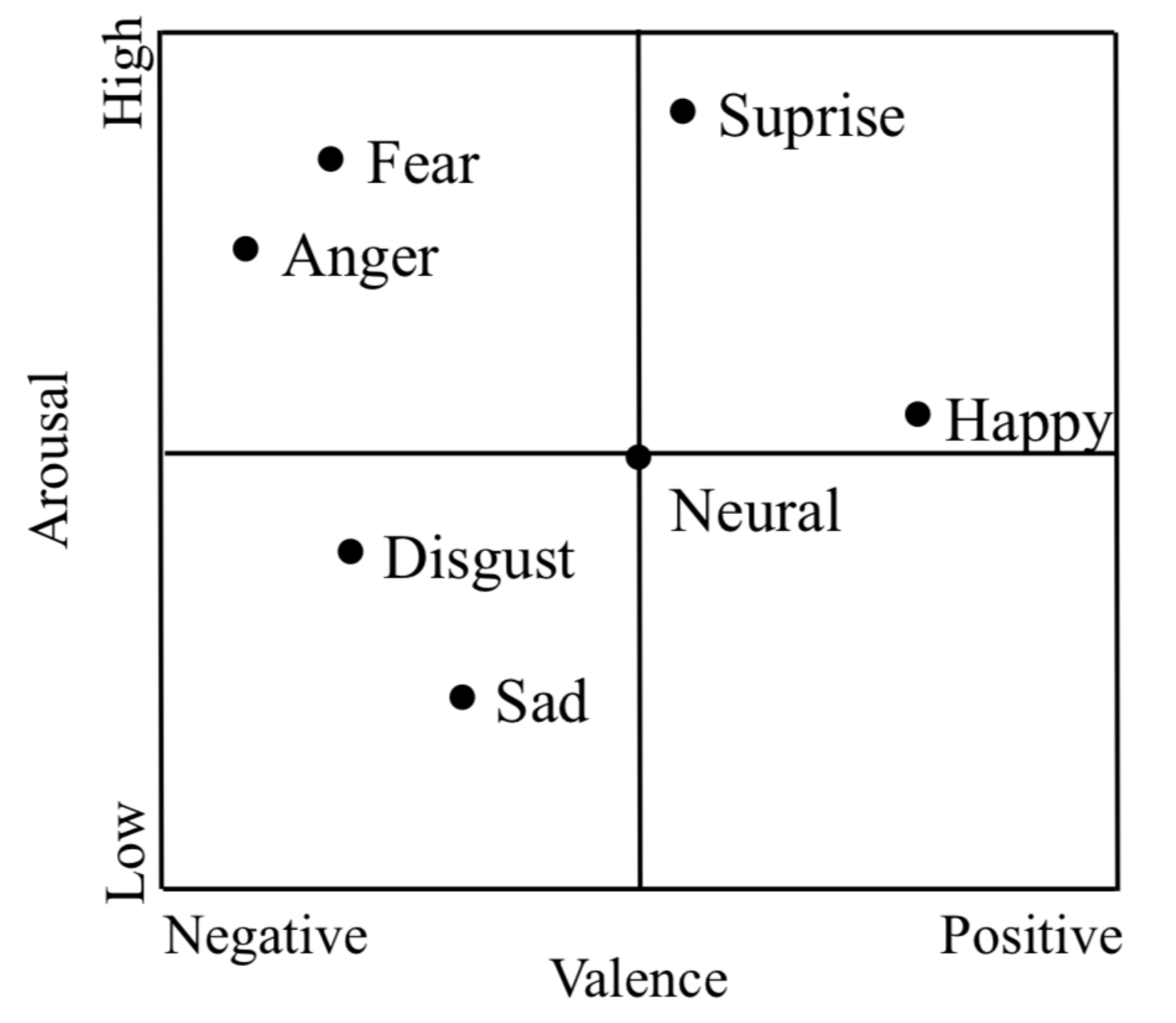}
     \caption{2D Model by Arousal and Valence}
     \label{fig:2D_Model}
     \end{subfigure}
     \hfill
\caption{Network Architectures} % Add some conclusions
\label{fig:net_archi}
\end{figure}

\textbf{Network Architecture.}
We try two network architectures in this paper, as shown in Figure~\ref{fig:net_archi}. The basic network is a simple three linear layers network, consisting of two hidden layers and an output layer. BatchNorm and ReLU is applied after each hidden layer. Softmax is employed as the activate function after the output layer. The number of hidden units in each hidden layer is 16. The output of the network is a 7-dimension vector. We also provide a two-step network, which use a three layer network to classify the signals into high/low arousal or positive/negative valence at the first stage according to the 2D model illustrated in Figure~\ref{fig:2D_Model}. The Sigmoid function is employed before the 1-dimension mid-output of the first classification network. Then the mid-output is concatenated with the features as the input of the second 7-classification network, which is the same as the basic network architecture.
Binary Cross Entropy loss is utilized as the training objective function for both architectures. 

\section{Experiments}
\label{sec:Experiments}
In this section, we conduct three groups of experiment. We first train and test with the 7 class emotion data to show the general quality of our dataset. Then we train a network to classify the data according to the 2D emotion model in Figure~\ref{fig:2D_Model} by arousal and valence separately. After that, we train the two-step network to verify whether there will be any improvement with the arousal/valence information added.

\subsection{Training details and measurement}
The networks are implemented with Pytorch 1.3.1 and trained on an Intel(R) Core(TM) i7-9700K processor with 3.60 GHz, 64.00 GB of RAM with Ubuntu 16.04 LTS installed. The labels for time series EDA data are given by the labels of the corresponding videos, as mentioned before. The classification accuracy is reported as measurement which shows the proportion of accurate prediction. Leave-one-observer-out is employed and the average accuracy of all the runs measures the classification accuracy of our baseline network. The learning rate we use is 1e-2 and we train for 700 epochs in all the experiment. % lr, epoches, average the results

\subsection{Results and discussion}
\textbf{The Basic Network.} We train and test on all the 7 class data and the results are listed in Table~\ref{Tab:Classification_Accuracy}. When we include the data from all participants, only single participants and only multiple participants in our training and test set, the average accuracy achieves 37.24\%, 42.08\% and 33.07\% respectively. The results illustrate that the data from single participants are more predictable than multiple participants. This might be because participating multiple times introduces more uncontrollable noise to the data, like participants may have different mood before watching the videos each time, or they are more bored later, and so on. We only use the single participant dataset in the following experiment.

\begin{table*}
\centering
\caption{Classification Accuracy}
\label{Tab:Classification_Accuracy}
\begin{tabular}{l|c|c|c}
\hline\hline
Group & All & Single & Multiple \\
\hline
Classification Accuracy  & 37.24\%  & \textbf{42.08\%} & 33.07\%  \\
% \multicolumn{2}{c|}{Without Group} & 30.62\%  & 37.77\% &  26.21\%  \\ \hline
% \multirow{2}{7em}{Group by Arousal} & High & 49.77\%  & 62.93\% & 40.12\%   \\
% & Low & 68.82\% & 74.21\% & 55.96\%   \\ \hline
% \multirow{2}{7em}{Group by Valence} & Positive & 53.60\%  & 60.20\% & 36.58\%  \\
% & Negative & 64.05\%  & 72.54\% & 73.69\%   \\
\hline\hline
\end{tabular}
\end{table*}

\textbf{The Classification Accuracy of Arousal and Valence.} From the vertical dimension in the 2D model shown in Fig.~\ref{fig:2D_Model}, the dataset can be split into 2 groups: a high arousal group consists of surprise, fear, anger and happy, which can be labeled as 1; a low arousal group which contains sad, neural and disgust, which can be labeled as 0. We employ the architecture of the basic network to classify the dataset into these two classes and the classification accuracy is \textbf{68.66\%}. Comparing with the random guess accuracy of 52.03\% of these two classes, our baseline architecture improves 16.63\%.
% Using single participants' data for training and testing obtains 62.93\% in high arousal group and 74.21\% in low arousal group. This illustrates that low arousal emotions are easier to predict than high arousal emotions. This difference is more than would be expected when predicting only 3 classes in low arousal group and 4 classes in high arousal group, \textcolor{red}{but also because} some emotions in the high arousal group like fear and surprise are hard to classify intrinsically.~\cite{}
From the horizontal dimension in the 2D model, the 7 emotions can be divided into a different 2 groups according to their valence: a positive group consists of surprise, neural and happy, which can be labeled as 1; a negative group which contains sad, anger, fear and disgust,which can be labeled as 0. The classification accuracy we achieve by the basic network is \textbf{72.72\%}, which is 22.10\% higher than the random guess accuracy 50.62\%.
This illustrates that classifying emotions into positive and negative according to valence is easier than classifying them into high and low arousal.
% Using single participants' data for training and testing obtains 60.20\% in positive group and 72.54\% in negative group. This demonstrates that negative emotions are easier to predict than positive emotions.

\textbf{The two-step network.} Then we train the two-step network. The loss function is a combination of the loss of the first 2-classification network and the loss of the 7-classification task. The mid-output by the first network, which indicates the arousal/valence information, is concatenated with the features as the input of the second network. We train this network end-to-end with the same hyper parameters as the basic network for a fair comparison. If we learn a classifier for arousal in the first network, the classification accuracy we achieve is \textbf{42.24\%} by this two-step network. If we learn a classifier for valence, the classification accuracy is \textbf{42.57\%}, which improves 4.9\% comparing with the basic network. Thus, adding arousal or valence information as an input can help to improve the classification accuracy. The information of positive/ negative valence boosts the classification accuracy to a higher degree on this dataset. This might be due to our finding that we can obtain a higher classification accuracy for valence comparing to arousal.

\section{Conclusion}
\label{sec:Conclusion}
This paper proposes a new physiological dataset PAFEW for observers' emotion classification. The dataset is collected by E4 wristband and Eye Tribe while asking subjects to observe the movie clips from AFEW~\cite{AFEW}. We extract 6 features from each data series corresponding to each video as the input of the network. Our three layer basic network achieves an accuracy of 42.08\% on the single participant dataset. We also train classifiers to classify the data by arousal and valence, resulting in an accuracy of 68.66\% and 72.72\% separately, which illustrates that classifying emotions into positive and negative according to valence is easier than classifying them into high and low arousal. Based on these results, we use a two-step network to add the arousal/valence information into the input features and achieve an improvement as far as 4.9\% comparing with the basic network. The information of positive/ negative valence brings a higher improvement to the classification accuracy of this dataset. 

\bibliographystyle{unsrt}  
\bibliography{references}  %%% Remove comment to use the external .bib file (using bibtex).
\end{document}